\documentclass{sf2a-conf2022}
\usepackage{graphicx}
\usepackage{hyperref}
\usepackage[]{natbib}  
\usepackage{epstopdf}

\def\BibTeX{{\rm B\kern-.05em{\sc i\kern-.025em b}\kern-.08em
    T\kern-.1667em\lower.7ex\hbox{E}\kern-.125emX}}
\bibpunct{(}{)}{;}{a}{}{,}  

\begin{document}

\TitreGlobal{SF2A 2022}


\title{Gender and precarity in Astronomy}

\runningtitle{Gender and precarity in Astronomy}

\author{N. A. Webb}\address{IRAP, Universit\'e de Toulouse, CNRS, CNES, Toulouse, France}

\author{C. Bot}\address{CDS, Universit\'e de Strasbourg, CNRS, Observatoire astronomique de Strasbourg, UMR 7550, 67000 Strasbourg, France}
\author{S. Charpinet$^1$}
\author{T. Contini$^1$}
\author{L. Jouve$^1$}
\author{H. Meheut$^3$}\address{Universit\'e C\^ote d'Azur, Observatoire de la C\^ote d'Azur, CNRS, Laboratoire Lagrange, Laboratoire Art\'mis, France}
\author{S. Mei}\address{Universit\'e de Paris, CNRS, Astroparticule et Cosmologie, F-75013 Paris, France}
\author{B. Mosser}\address{LESIA, Observatoire de Paris, PSL Research University, CNRS, Universit\'e Pierre et Marie Curie, Universit\'e Paris Diderot, 92195 Meudon, France}
\author{G. Soucail$^1$}






\setcounter{page}{237}


\maketitle


\begin{abstract}
  Following the survey {\em Well-being in astrophysics} that was sent out in March 2021, to establish how astrophysics researchers, primarily in France, experience their career, some of the results were published in \cite{webb21}. Here we further analyse the data to determine if gender can cause different experiences in astrophysics. We also study the impact on the well-being of temporary staff (primarily PhD students and postdocs), compared to permanent staff. Whilst more temporary staff stated that they felt permanently overwhelmed than permanent staff, the experiences in astrophysics for the different genders were in general very similar, except in one area. More than three times more females than males experienced harassment or discrimination, rising sharply for gender discrimination and sexual harassment, where all of those having experienced sexual harassment and who had provided their gender in the survey, were female. Further, as previously reported \citep{webb21}, 20\% of the respondents had suffered mental health issues before starting their career in astrophysics. We found that whilst this group was split approximately equally with regards to males and females, the number rose sharply to almost 45\% of astronomers experiencing mental health issues since starting in astrophysics. Of this population, there were 50\% more females than males. This excess of females was almost entirely made up of the population of women that had been harassed or discriminated against.

\end{abstract}



\section{Introduction}

In order to establish how astrophysics researchers, primarily in France, experience their career, we sent out a survey in March 2021, to understand the impact that their job was having on their well-being. 276 people responded to the survey and some initial results were published towards the end of the year \citep{webb21}.  Amongst other things in \cite{webb21}, we showed that around half of the respondents expressed pleasure derived from their career, but it was also clear that many (early career) researchers were suffering due to overwork, with more than a quarter saying that they work in excess of 50 hours per week and 2\% in excess  of 90 h per week. Almost 30\% professed to having suffered harrassment or discrimination in the course of their work. Further, whilst only 20\% had suffered mental health issues before starting their career in astrophysics, $\sim$45\% said that they suffered with mental health problems since starting in astrophysics.

In order to understand the origin of the mental health problems and explore in greater detail whether there was any difference depending on gender or between temporary and permanent staff, we examined the data further and the results on these aspects are presented in the following.

\section{Comparison of temporary and permanent staff}

The majority of respondents to the survey were temporary staff (undergraduate or masters students, PhD students or post-doctoral scholars), as the survey was aimed mostly at this population, as many studies \citep[e.g.][]{auer18,wool19} had already shown that this population was experiencing the most difficulties. This group amounted to 215 people and 61 permanent staff responded to the survey. Correcting for the different population sizes as we do for gender, see Section \ref{sec:gender}, it was clear that the majority of people responding to the survey that said that they had experienced harassment/discrimination was in large majority the post-docs.  Concerning colleagues feeling overwhelmed at work, see Fig.~\ref{fig:overwhelmed}, those that felt overwhelmed all of the time were {\it all} temporary staff.

\section{Comparison of the genders}
\label{sec:gender}

For those people that provided their gender, 57.8\% identified as male, 40.6\% as female and 1.6\% identified as other. As few people identified as other, it is difficult to obtain reliable conclusions for this group and therefore we focus mainly on the male and female groups in this work. The data presented in the following is corrected for the differences in the number of women and men that answered the survey, so the percentages provided correspond to a sample of 50\% male and 50\% female. In the majority of the areas we addressed, namely people's favourite aspects of astrophysics research, the number of hours worked, perceived external constraints, job satisfaction and future plans for remaining in the field, the results were very similar for both males and females.

There were slight differences in living conditions for males and females. 17\% of respondents said that they have children to care for and of these, 63\% were male. Conversely, 7\% stated that they have other family members to care for, and here 61\% were female. 25\% replied saying that they live apart from their partner and/or children and slightly more, 61\% were female. 12\% of respondents admitted to having debt problems, 66\% being female. 

However, there were a couple of areas where the differences between male and female respondants were highly different and that was in harassment/discrimination and health. Overall, 3.3 times more women experienced harassment/discrimination, than men.  Figure 2 in \cite{webb21} showed the number of people that had experienced different forms of harassment or discrimination. In Table~\ref{tab:harassment} we provide the type of harassment/discrimination when at least five people indicated that they had been subjected to these problems. We also provide the total number of people having been subjected to this type of harassment/discrimination, and the ratio of women to men that were the subject of these issues.

\begin{table}
      \caption[]{Types of harassment/discrimination experienced by at least five respondents, number of people subjected to the issue and the ratio of women to men subjected to the problem, for which the gender was known.}
         \label{tab:harassment}
\begin{center}
   \begin{tabular}{lcc}     
\hline\hline       
Harassment/discrimination & Number of people subjected to issue & Ratio of women to men \\
\hline                    
Sexual harassment & 15 & $\infty$\\
Age discrimination & 21 & 4.5 \\
Racial discrimination/harassment & 6 & 7 \\
Gender discrimination & 39 & 22\\
Shaming by a superior or a colleague & 33 & 3.1 \\
Harassment by supervisor/superior & 29 & 3.2\\
\hline
\end{tabular}
\end{center}
   \end{table}  

\begin{figure}[ht!]
 \centering
 \includegraphics[width=0.7\textwidth,clip]{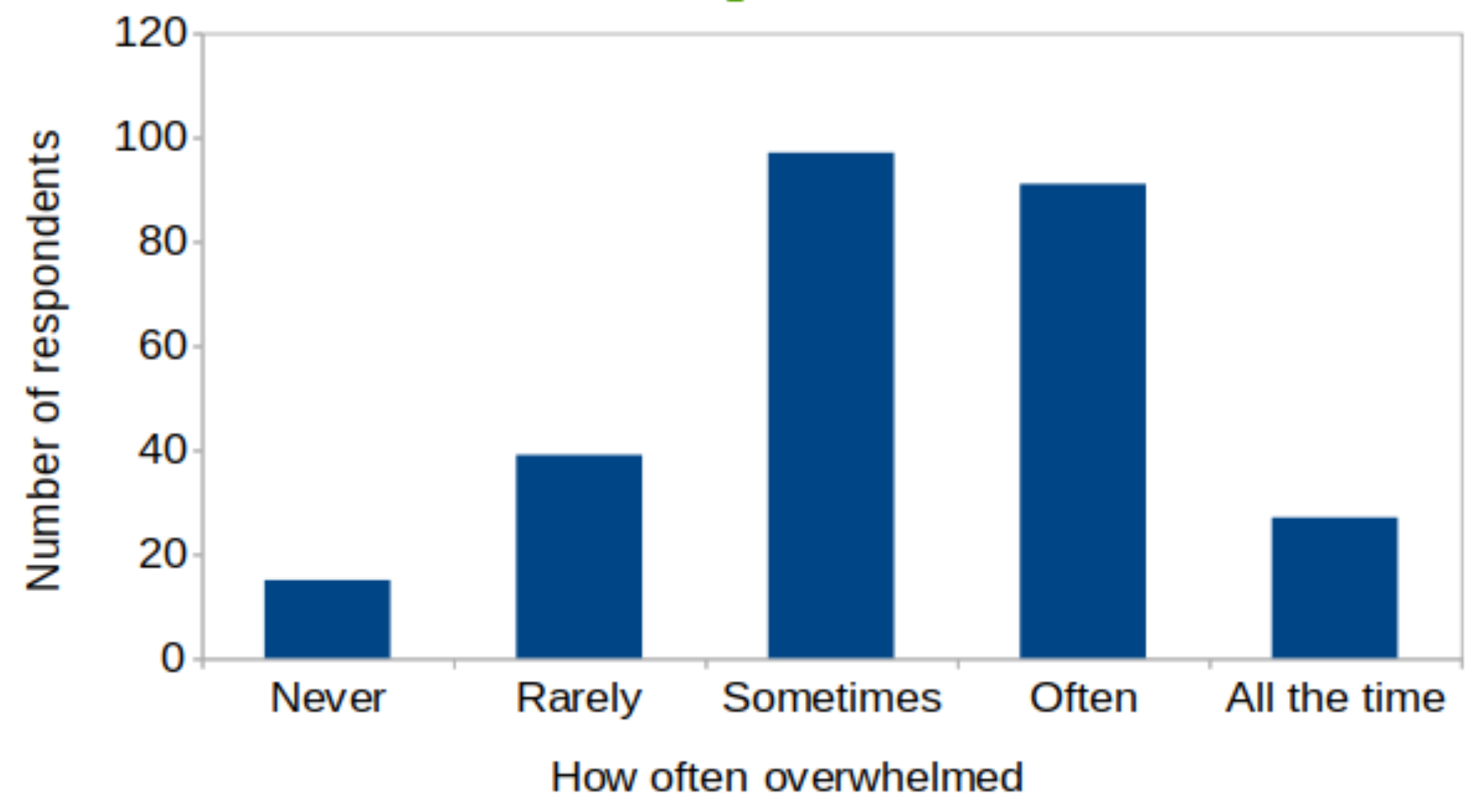}      
  \caption{Number of respondents that felt overwhelmed never, rarely, sometimes, often or all the time.}
  \label{fig:overwhelmed}
\end{figure}

With regards to health issues, only a third of all respondents said that they were happy with their health, with significantly less than half of these being female.  Figure~\ref{fig:overwhelmed} shows how often respondents felt overwhelmed. The histogram shows the number of respondents for the five possibilities, never, rarely, sometimes, often or all the time.  For the group that felt overwhelmed either {\it often} or {\it all of the time}, there were 50\% more females than males.  44\% of all respondents felt that the expectations were too high in astrophysics and of these, there were again 50\% more females than males. Only 37\% of those that felt that senior colleagues were there to provide support when they needed it, were female.

As stated above, 20\% of the respondents had suffered mental health issues before starting their career in astrophysics and these were approximately equal in proportion between males and females. Since starting in astrophysics, $\sim$45\% suffered with mental health problems. Of this population, there were 50\% more females than males. The number of people suffering with bipolar disorder, eating/body dismorphic disorder, or obsessive-compulsive disorder remained the same, but for depression, anxiety and panic attacks, the number of people more than doubled in each case. Whilst our survey took place since the beginning of the COVID-19 pandemic and some of the cases could be attributed to the restrictions introduced to stem infections spreading, the majority of people that started suffering with mental health issues since starting astrophysics, is not only women, but are those that have been harassed or discriminated against.

\section{Discussion}

It is clear from the survey that colleagues on temporary contracts are frequently overwhelmed. This may be due to the workload, down to high expectations or due to the short contracts, where a great deal is expected from a PhD student or a post-doc in a short period of time. This can be exacerbated by those working in projects funded through the {\em Agence Nationale de la Recherche} (the French National Research Agency, ANR) or the {\em European Research Council} (ERC), for example, or in big collaborations, as deadlines to deliver are often frequent and difficult to achieve. However, aspects external to the astronomy part of the work can also contribute to temporary contract colleagues feeling overwhelmed, including: living away from family, be it a partner, children or parents, which can also remove a support network; cultural and administrative problems i.e. with paperwork for the job or a visa, a possible language barrier in a new country, insufficient information on institute practices, etc; or financial difficulties due to months between contracts or due to moving to a new institute/country. This may also be due to harassment or bullying and this is discussed further below.

More than three times as many women than men responding to the survey have experienced harassment or discrimination in astrophysics in France. Women and men experienced all types of harassment and discrimination. However, women were the large majority in gender discrimination, where of the 39 people that responded to say that they had experienced this kind of discrimination, 32 were female, 2 were male and 0 were 'other'. Five people did not give their gender. Worse still was sexual harassment, where all of those that experienced these issues and who provided their gender were female.  It is clear that we are still a long way from gender equality in the world of astrophysics.

The negative outcome of this harassment is that it has lead to mental health issues, including depression, anxiety and panic attacks, making people's daily lives miserable. This is not the first survey to reveal such facts. A recent survey\footnote{\url{https://ras.ac.uk/news-and-press/news/survey-finds-bullying-and-harassment-systemic-astronomy-and-geophysics}} of astronomers and geophysicists commissioned by the {\it  RAS Committee on Diversity in Astronomy and Geophysics } and that ran in the UK a year prior to our survey, but for which the results are only starting to appear, found that 44\% of respondents had suffered bullying and harassment in the workplace within the preceding 12 months. They also stated that ``Women and non-binary people in the field are 50\% more likely than men to be bullied and harassed.'', exactly in line with our findings. Thanks to more than twice the number of respondents in the UK survey compared to the French survey (650 respondents from the UK astronomy and geophysics communities), they were also able to draw conclusions on other minorities. They found that 50\% of lesbian, gay, bisexual, and queer astronomers and geophysicists were bullied in the 12 months prior to the survey, and that 12\% of bisexual astronomers reported being bullied at least once a week!

The {\it American Institute of Physics} (AIP) carries out the Longitudinal Survey of Astronomy Graduate Students\footnote{\url{https://www.newswise.com/articles/aip-report-harassment-discrimination-in-astronomy-takes-many-forms}}. It was initiated by the {\it American Astronomical Society} (AAS) in 2006 to gain an understanding of the forms and long-term impacts of harassment in astronomy. This study was carried out by polling astronomy graduates from 2006-07 and followed them in both 2012-13 and in 2015-16 after they entered the workplace. This study showed that 33\% of the respondents had experienced harassment and discrimination whilst studying or working. The AIP broke the harassment/discrimination that was reported into four general groups. These were: biased assumptions (regarding status, career, or personal life); verbal put-downs (jokes, criticisms, or undermining comments); demographic-based inequitable treatment, limiting social support or professional development; and unwanted sexual attention, ranging from inappropriate comments to more serious behaviors such as threats, stalking, and assault. The AAS has tried to improve things by introducing a {\it Code of Ethics} and an {\it Anti-Harassment Policy} for AAS-sponsored meetings, publications, etc and supported by a complaint process and an Ethics Committee. They plan to implement a new survey to determine whether these policies have helped improve the situation.

Pending the results of the new AAS survey and thus understanding the effectiveness of the measures implemented by the AAS and whether they should be implemented more broadly, more can still be done. This could be through education, as some perpetrators of harassment or discrimination may not necessarily be aware that their actions are having a negative impact. For example, long working hours have long been endemic in astronomy and some researchers may think that this is the only way astronomy can work, and insist that colleagues work very long hours, as they do not know otherwise. Also, in the past, astronomy has lacked diversity and some colleagues may lack experience in interacting with people from other backgrounds. Whilst it is clear we should all be respectful of each other, it may happen that colleagues inadvertently say something inappropriate.  Becoming familiar with the range of diversity, be it gender, race, neurodiversity or other, can help to avoid inappropriate behaviour. As outlined in \cite{webb21}, everyone working in astrophysics should be trained in well-being, harassment, discrimination etc in order to be aware of diversity and be familiar with the language associated with these issues. We recommend using online courses with tests, which need to be passed every few years or through information provided in each institute.

Many people in academia feel a sense of shame when they recognise that they have been the target of negative behaviour \citep{keas19}. This can extend to feelings of personal failure, or put them at risk for more targeting.  As many as 20-40\% of academic personnel have been harassed, and 50\% have witnessed it according to \cite{keas19}. It is therefore essential that anyone is able to report cases of harassment or discrimination. Action has been taken to ensure that every astrophysics laboratory in France has a person to whom such problems can be reported\footnote{\url{https://www.dgdr.cnrs.fr/bo/2021/BO-avril-2021-compressed.pdf}}. However, care must be taken as our survey showed that in many cases, people did not report the harassment. Only 26\% of respondents that experienced such issues said that they had reported them. Of those that did not report the harassment, 40\% did not report it as they felt that they wouldn't be listened to. A further 25\% did not report the harassment, as they felt that other people have worse problems and that it was not worth reporting. 18\% were embarassed and 12\% did not even know where to go to report the problem. Finally 5\% did not report the problem due to not speaking the local language. This is a major hindrance to helping with the problem, but it appears that this culture of silence is changing, though this change is still slow \citep{keas19}. To ensure that all problems are reported, so that action can be taken, it is important that everyone feels that they are listened to, no matter how big or small the problem and have people that are able to speak both the local language and English, which is widely spoken by researchers in astronomy. Having an email address where people can send a short report of a seemingly minor incident and the name of the perpertrator, could help to highlight a perpertrator that carries out micro-agressions that may go unnoticed as isolated incidents, but that could be frequently carried out on a number of different people. However, the modality for putting this into place should be discussed to avoid misuse. Finally, it is essential that if colleagues witness harassment or bullying that they report the incident or intervene, if they feel that they are able to.

Other areas that can be explored to improve the situation are introducing Codes of Conduct in different organisations, in a similar way to the work done by the AAS, specifying clearly what behaviour is acceptable. Codes of Conduct are becoming more common, so it should be possible to see an improvement in the coming years. Positive conflict management practices should be introduced. They should include clear procedures and dedicated committees and ombudspersons. Through listening and searching for common ground, collaborative solutions should be found. Some collaborations and organisations have now trained teams (allies) to observe interactions and help identify improvements which should also help improve situations. Finally training and dedicated talks to raise awareness should be held. These should provide the background needed for everyone to participate and create a positive environment.

In France improvements are being introduced. In addition to the action taken to ensure that every astrophysics laboratory has a person where harassment and discrimination can be reported, a group associated with the National society for astronomy and astrophysics (SF2A) has been put into place and national workshops to discuss the problems have been held. A well-being webpage on the SF2A website is being put into place to centralise information that is already available, but maybe difficult to find. It includes a description of different issues and examples of problems to make it clear to everyone, as well as where to find help. Meetings with gouverning bodies for astrophysics have been held to try to ensure that paperwork is always provided in the local language {\bf and} in English in every institute and in institutes where it is not yet done, up to date, introductory information about the institue should be provided to new arrivals either through welcome guides or through bi-annual meetings (or both). It is also recommended that more (non-)scientific events are organised to encourage a team spirit and more open discussion on well-being, working hours, racism, sexism, etc should be organised. The possibility of harmonising post-doc salaries on a national level and provide an evolution in the renumeration is being investigated. Many other actions are taking place on a local level in different French institutes.

\section{Lessons learnt}

This survey was a first attempt to gain insight into the current state of well-being of primarily French astronomers. It has been highly instructive in showing that a large proportion of our colleagues are suffering due to the environment in which they work. It has also provided suggestions as to how to make improvements. A second survey is planned within a year to extend the participation to include all colleagues (young and old) across physics. We are also investigating the possibilty of extending the survey to other countries and to make the survey more inclusive to include engineers working in the domain who are also affected by similar problems. This will increase the number of participants and allow us to improve statistics to better understand the well-being of minority groups and make a fairer comparison of colleagues in different age ranges and positions. Allowing two years between the surveys will allow us to ascertain if any of the actions we have started to put into place are indeed helping to improve well-being.

\section{Conclusions}

It is clear that we are still a very long way from having equality in astrophysics. Whilst with the data available in our survey we could only investigate the prevelance and impact of harassment and discrimination on two main groups, male and female, other studies have shown that minorities other than women are also disproportionately impacted by harassment and discrimination and that this is having a major impact on mental health. It is clear that this has to stop and through education and speaking out about the problem, the situation can be improved. We invite you all to be a part of making astronomy a safer and a happier environment to work in, which will also help to make it a more productive and prolific one.

\bibliographystyle{aa}  
\bibliography{webb1} 

\begin{thebibliography}{4}
\expandafter\ifx\csname natexlab\endcsname\relax\def\natexlab#1{#1}\fi

\bibitem[{{Auerbach} {et~al.}(2018){Auerbach}, {Mortier}, {Bruffaerts},
  {Alonso}, {Benjet}, {Cuijpers}, {Demyttenaere}, {Ebert}, {Green}, {Hasking},
  {Murray}, {Nock}, {Pinder-Amaker}, {Sampson}, {Stein}, {Vilagut},
  {Zaslavsky}, {Kessler}, \& {WHO WMH-ICS Collaborators}}]{auer18}
{Auerbach}, R.~P., {Mortier}, P., {Bruffaerts}, R., {et~al.} 2018, Journal of
  Abnormal Psychology, 127, 623

\bibitem[{Keashly(2019)}]{keas19}
Keashly, L. 2019, Workplace Bullying, Mobbing and Harassment in Academe:
  Faculty Experience, ed. P.~D'Cruz, E.~Noronha, L.~Keashly, \& S.~Tye-Williams
  (Singapore: Springer Singapore), 1--77

\bibitem[{{Webb} {et~al.}(2021){Webb}, {Bot}, {Charpinet}, {Contini}, {Jouve},
  {Koliopanos}, {Lamberts}, {Meheut}, {Mei}, {Ristorcelli}, \&
  {Soucail}}]{webb21}
{Webb}, N.~A., {Bot}, C., {Charpinet}, S., {et~al.} 2021, in SF2A-2021:
  Proceedings of the Annual meeting of the French Society of Astronomy and
  Astrophysics, ed. A.~{Siebert}, K.~{Bailli{\'e}}, E.~{Lagadec}, N.~{Lagarde},
  J.~{Malzac}, J.~B. {Marquette}, M.~{N'Diaye}, J.~{Richard}, \& O.~{Venot},
  35--40

\bibitem[{{Woolston}(2019)}]{wool19}
{Woolston}, C. 2019, Nature, 575, 403

\end{thebibliography}

\end{document}